\documentclass[paper,twocolumn,superscriptaddress,apl]{revtex4}%

\usepackage{amsfonts}
\usepackage{amsmath}
\usepackage{amssymb}
\usepackage{graphicx}
\usepackage{color}

\begin{document}

\title{Robust data analysis and imaging with computational ghost imaging}


\author{Jiangtao Liu}
\email[Electronic address:]{jtliu@semi.ac.cn}
\affiliation{College of Mechanical and Electrical Engineering, Guizhou Minzu University, Guiyang 550025, China}

\author{Xunming Cai}
\affiliation{College of Mechanical and Electrical Engineering, Guizhou Minzu University, Guiyang 550025, China}

\author{Jinbao Huang}
\affiliation{College of Mechanical and Electrical Engineering, Guizhou Minzu University, Guiyang 550025, China}

\author{Kun Luo}
\affiliation
{Key Laboratory of Microelectronic Devices and Integrated Technology, Institute of Microelectronics, Chinese Academy of Sciences, Beijing 100029, China}

\author{Hongxu Li}
\affiliation{Guizhou Light Industry Technical College, Guiyang 550025, China}

\author{Dejian Zhang}
\email[Electronic address:]{dejianzhang@ncu.edu.cn}
\affiliation{Nanchang University, Department of Physics, Nanchang, China}

\author{Zhenhua Wu}
\email[Electronic address:]{wuzhenhua@ime.ac.cn}
\affiliation
{Key Laboratory of Microelectronic Devices and Integrated Technology, Institute of Microelectronics, Chinese Academy of Sciences, Beijing 100029, China}
\affiliation{University of CAS, Beijing 100049, China}


\begin{abstract}
Nowadays the world has entered into the digital age, in which the data analysis and visualization have become more and more important. In analogy to imaging the real object, we demonstrate that the computational ghost imaging can image the digital data to show their characteristics, such as periodicity. Furthermore, our experimental results show that the use of optical imaging methods to analyse data exhibits unique advantages, especially in anti-interference. The data analysis with computational ghost imaging  can be well performed against strong noise, random amplitude and phase changes in the binarized signals. Such robust data data analysis and imaging has an important application prospect in big data analysis, meteorology, astronomy, economics and many other fields. \end{abstract}

\maketitle

The emergence of equipment and technologies for the optical imaging of real objects such as telescopes and microscopes has led to rapid developments in astronomy and biology. However, we have entered the digital age, and the development of information technology,  has made data analysis and visualization increasingly important. Additionally, the exponential growth of massive amounts of data makes data analysis increasingly difficult, and new data analysis and visualization technologies are urgently needed. In view of the great success of the optical imaging of physical objects, imaging data with optical methods to obtain the characteristics of the data would be an interesting challenge.

In this study, we found that ghost imaging can not only image a physical object but also image data, presenting the characteristics of the data. Ghost imaging is a new imaging technology based on the correlation function between a reference light field and the target detection light field\cite{ZETF90KDN,ZETF78BAV,NP13MPE,LPR18MPA}. Ghost imaging has important applications in 3D imaging\cite{S340SB}, imaging with atoms\cite{N540RIK,PRL121LS}, time domain imaging\cite{NP10PR,o6HW}, X-ray imaging\cite{PRL92CJ,o5axz},
Neutron  imaging\cite{PRA101KA}, THz wave imaging\cite{LSA9SCC,O7LO}, super-resolution imaging\cite{O6WL}, thermal light imaging\cite{PRL94VA,ol34xhc,APL117YZ,PRA15ZDJ}, difference imaging\cite{PRA15YZ}, helicity-dependent metasurface imaging\cite{SA3LH}, ultrafast pump-probe ghost imaging\cite{PRX9RD,APL116WX}, edge imaging\cite{APL116SH},  optical encryption\cite{OL35PC}, etc.  Ghost imaging is a nonlocal imaging method that is robust against interference and penetrates scattering media\cite{NP13MPE,LPR18MPA,OE17JC,APL98MRE}.

Specifically, we use computational ghost imaging to image the periodicity in data. Periodicity is one of the most important and basic trends. As early as 1705, Halley successfully predicted the reappearance of Halley's comet by analyzing the periodicity of the appearance of the comet. Today, periodic analysis is also used to study the hidden trends of climate change \cite{N367MES,S268GP}, earthquakes \cite{NSR6JH}, astronomy \cite{ND89SVG,TAJ837BS,TAJ883BDN}, infectious diseases \cite{PNAS99RX}, biology \cite{NAR46ZX,SR8HS}, oceanography \cite{PNAS109GD} and so on. At present, periodic analysis mainly includes singular spectrum analysis\cite{BPP1996EJB}, fast Fourier transform\cite{MC19JWC}, uncoiled random QR denoising\cite{PNAS111CL}, deviance information criterion minimization \cite{TAJ853KT}, etc. However, in the digital age, due to the diversification of the data collected, the accuracy of the data decreases, and the hybridity continues to improve. There is substantial noise in too large of a dataset, and imprecise acquisition makes the amplitude and phase of the collected periodic signal highly random. In addition, to reduce the information storage and improve the analysis speed, many data are binarized. These characteristics mean that traditional periodic analysis methods are no longer applicable.

In our experiment applying ghost imaging to data, we find that random factors such as noise, random amplitudes and phase changes in the data mainly affect the object
imaging area the but have little influence on the background imaging and that the periodicity in the data is clearer in the background imaging area.
Thus, the signal and noise can be separated, and the influence of the signal amplitude and phase randomness caused by inaccurate acquisition and other factors can be eliminated.
Moreover, in computational ghost imaging, most speckle patterns are binary, so they are compatible with binary data. Our experimental results show that the periodicity of binary data can be
presented when the signal-to-noise ratio (SNR) is 1:19 or when the SNR is 1:7, the phase random amplitude reaches $210^\circ$, and the amplitude random amplitude reaches 90\%.

\begin{figure}[t]
 \centering\includegraphics[width=0.9\columnwidth,clip]{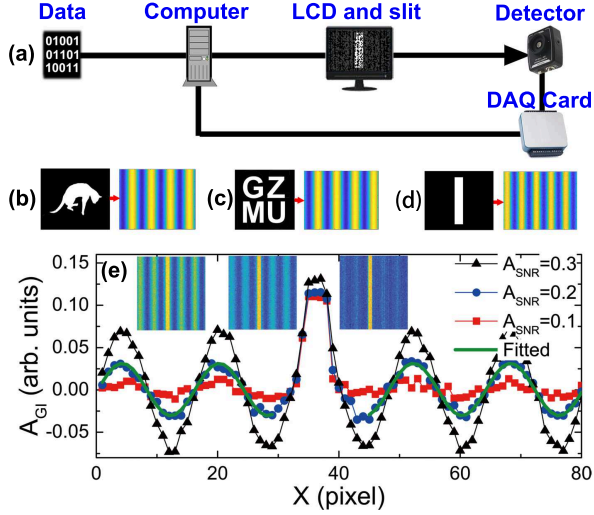}
\caption{Schematic diagram of a lensless ghost imaging system based on a liquid crystal display. (b)-(d) Different objects to be imaged and the corresponding images under periodic signals, the period of the signal in (d) is different. The brightness distribution of the image with different periodic signal amplitudes $A_{SNR}$, where the left, middle and right insets are $A_{SNR}=0.3$, $A_{SNR}=0.2$, and $A_{SNR}=0.1$, and the green solid line fits the curve with $A_{SNR}=0.3$.
}
\label{fig1}%
\end{figure}

To eliminate the image distortion caused by a lens, we build a lensless ghost imaging system using a ACER  XB252Q liquid crystal display [Fig. 1 (a)]. The data to be analyzed are used to generate the corresponding speckle pattern, and the light emitted by the speckle pattern passes through the object to be imaged and is received by the Thorlabs
PDA36A2 light detector, and the signals is collected by a NI USB 6002 data acquisition card, and the corresponding image is obtained by the correlation calculation of the signal received by the bucket detector and the speckle pattern. If the data to be analyzed contain periodic signals, periodic stripes-shaped images are obtained [Figs. 1 (b) and (c)]. The image shows the characteristics of the data independent of the shape of the object to be imprinted, and the periodicity of the stripes is exactly the same as that of the data. When the period of the signal changes, the period of the image will also change [Fig. 1 (d)].

The main reason for this is that the computed ghost image is obtained by association calculation between the received signal of the bucket detector $I_1$ and the corresponding incident speckle field $I_2(x,y)$, i.e.,
\begin{equation}
\begin{aligned}
\Delta G(x,y)=&\langle (I_{1}-\langle I_{1}\rangle) [I_{2}(x,y)-\langle I_{2}(x,y) \rangle] \rangle \\
                   =&\langle I_{1}I_{2}(x,y)-\langle I_{1}\rangle\langle I_{2}(x,y) \rangle \rangle,
\label{Gcal:eq1}%
\end{aligned}
\end{equation}
where $\left< \cdots \right>$ denotes the ensemble average over thousands of measurements. Speckle patterns are generated by periodic signals such as $I_{2}(x,y)=\{ 1+\cos[2\pi(x_{R0}+x)/P_{d}] \}/2$, where $x_{R0}$ is a random initial pixel position and $P_{d}$ is the period length. Suppose there is a slit at $x=x_{0}$; thus, it is only around $x_{R0}+x_{0}=n P_{d}$ that $I_{1}$ does not equal 0, so we can obtain $\Delta G(x,y)\propto 1+\cos[2\pi(n P_{d}-x_{0}+x)/P_{d}] $. That is, the image shows the same periodicity as the speckle pattern.

However,  the actual periodic signal is often accompanied by a strong noise signal, and the periodic signal itself often exhibits random amplitude and phase changes. To better analyze the influence of periodic signal characteristics and noise on the calculation of ghost imaging, we  construct a binary database containing one-dimensional periodic signals with noise signal, random amplitude and random phase, which can be written as
\begin{equation}
\Lambda(x_{ip})=A_{SNR}[1+P(x_{ip})]/2+(1-A_{SNR})\mathcal{R}_{0,1},
\label{Rdata:eq2}%
\end{equation}
where $P(x_{ip})=\mathcal{R}_{A}\cos(2\pi x_{ip}/P_{d}+\mathcal{R}_{\varphi})$ is a periodic signal, $\mathcal{R}_{A}=(1-A_{RAC})\mathcal{R}_{0,1}+A_{RAC}$, $\mathcal{R}_{\varphi}=\pi A_{RPC}\mathcal{R}_{0,1}/180^{\circ}$, $\mathcal{R}_{0,1}$ is a uniformly distributed random number from 0 to 1, $A_{SNR}$ is the periodic signal amplitude, and $\mathcal{R}_{A}$ and $\mathcal{R}_{\varphi}$ reflect the randomness of the amplitude and phase, respectively. In addition, to make the imaging smoother, we add a two-dimensional random modulation; that is, $\Lambda'(x_{ip},y_{ip})=0.5\Lambda(x_{ip})+0.5\mathcal{R}_{0,1}(x_{ip},y_{ip})$. Therefore, when $A_{SNR}=0.1$, the corresponding signal (noise) amplitude is 0.05 (0.95), and the SNR is 1:19. We randomly obtain a segment from this database and binarize it to generate the corresponding speckle image and use it to achieve ghost imaging. Unless otherwise specified, we take $P_{d}=16$ during the imaging process, and the speckle pattern has a size of $240\times240$ pixels. Each speckle has a size of $4\times4$ pixels, and the number of speckle images is 3000. The object to be imaged is a slit with a width of 4.3 mm.

We study the influence of noise on periodic signal acquisition. When the signal amplitude $A_{SNR}=0.3$, that is, the SNR is 3:17, the generated image presents clear periodic stripes [the left inset of Fig. 1(e)]. However, compared with the noiseless periodic signal [Fig. 1(d)], the brightness of the middle fringe (i.e., the image corresponding to the object to be imaged) increases, while the brightness of the neighboring fringe decreases and the contrast decreases. As shown in traditional computed ghost imaging, the noise part of the speckle image images of the object, which increases the brightness of the image corresponding to the object. When the signal amplitude further decreases and the noise signal increases, such as $A_{SNR}=0.2$ or $A_{SNR}=0.1$ (SNR is 1:19), the contrast of the stripes is reduced, and the brightness of the middle stripes is increased, but the periodicity of the fringe can still be observed by the naked eye.

To better display the periodicity in the image, we sum the brightness of each pixel in the image along the vertical direction to obtain the brightness distribution of the stripes [as shown in Fig. 1(e)]. Then, curve fitting is performed on the area outside the middle bright stripe, and the error of the periodic signal extracted by ghost imaging can be obtained, e.g.,  when  $A_{SNR}=0.3$, the period (relative amplitude) errors are 0\% (1\%).

Next, we study the influence of the amplitude randomness and phase randomness of periodic signals on the acquisition of periodic signals. When $A_{RAC}$ changes from 0.5 to 0, that is, the signal amplitude randomness changes from 50\% to 100\%, the contrast of the fringe decreases slightly, but the generated images all show clear periodic stripes and do not show the image of the object to be measured [Fig. 2(a)]. The randomness of the amplitude of the periodic signal leads to a very small error.  However, in traditional periodic signal extraction, such as with the least squares method, the randomness of periodic signal amplitude increases the difficulty of signal extraction. When the phase is random, the resulting image presents periodic stripes and the image of the object. The effect of phase randomness on the image is similar to that of noise. When the phase randomness increases, the contrast of the fringe decreases, and the middle fringe increases the image brightness of the object. When $A_{RPC}$ is 210$^\circ$ and 300$^\circ$, the period error is approximately 0.1\%, and the relative amplitude error values are 1.2\% and 5.4\%, respectively. The brightness distribution curve of the image when noise, random amplitudes and random phases exist at the same time is shown in Fig. 2 (c). When $A_{SNR}=0.35$, $A_{RPC}=150^\circ$ and $A_{RAC}=0.3$ ($A_{SNR}=0.25$, $A_{RPC}=210^\circ$, and $A_{RAC}=0.1$,), the period error is -0.2\% ( -2.3\%), and the relative amplitude error is 2.1\% (9.6\%). By comparing the above results, we find that only when noise, random amplitudes and random phases are all present and large can a small periodic error ($\sim2.3\%$) occur, while in other cases, the periodic error is very small, so this method is excellent for periodic extraction in the signal.

\begin{figure}[t]
 \centering \includegraphics[width=0.9\columnwidth]{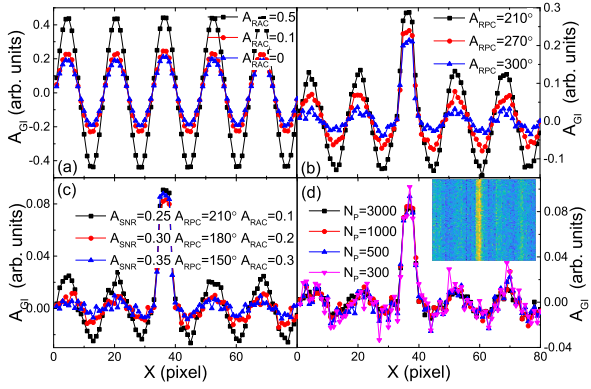}
\caption{(a) Periodic signal amplitude randomness and (b) phase randomness effects on the image brightness distribution. (c) The brightness distribution curve of the image when noise, random amplitude and random phase occur at the same time. (d) The influence of the number of speckle images on the image brightness distribution. The inset in (d) is the image generated when the number of speckle patterns is 300.
}
\label{fig2}%
\end{figure}

In addition, ghost imaging is insensitive to the number of speckle images in data imaging. Fig. 2 (d) shows the brightness distributions of images with different numbers of speckle images when $A_{SNR}=0.3$, $A_{RPC}=180^\circ$, and $A_{RAC}=0.2$. When the number of speckle images is 300, a clear periodic distribution can be presented. When the numbers of speckle patterns are 3000, 500 and 300, the corresponding periodic errors (the relative errors of amplitude) are $-0.9\%$ (6.1\%), $-0.3\%$ (9.0\%), and $1.1\%$ (10\%), respectively. This shows that this method can be used in data analysis with both small samples and large samples, so it has good flexibility.  In contrast, traditional ghost imaging generally requires a higher number of speckle images for physical imaging, and this defect limits the application of traditional ghost imaging in physical imaging. Furthermore, in the era of big data, the data itself is huge and fragmented, and each speckle pattern used in ghost imaging analysis of signals can be generated from a piece of fragmented information, and the number of speckle patterns can be more or less [Figure 2 (d)]. Therefore, ghost imaging can analyze both small fragmented data and huge fragmented data.  In addition, the number of calculations in this method is proportional to the amount of data, while the number of calculations in traditional methods such as singular spectrum analysis is proportional to the square of the amount of data. When the amount of data is large, this method has an advantage in calculation speed.

\begin{figure}[t]
 \centering \includegraphics[width=0.9\columnwidth,clip]{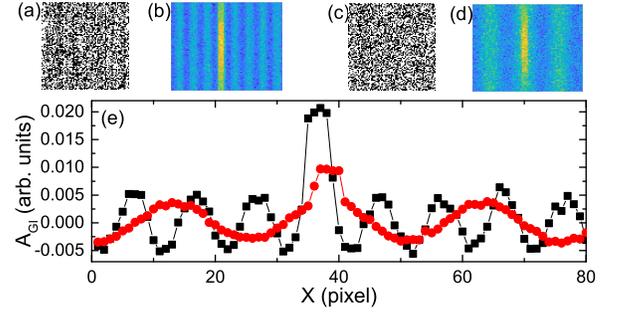}
\caption{(a) The speckle pattern generated by the actual circuit signal and (b) the resulting image  when SNR is approximately 1:9; (c) The speckle pattern generated by the actual circuit signal and (d) the resulting image  when SNR is approximately 1:2  and the random phase amplitude is approximately 290 degrees;  (d)Imaging brightness distribution, black line for SNR is approximately 1:2, red line for SNR is approximately 1:2  and the random phase amplitude is approximately 290 degrees.
}
\label{fig3}%
\end{figure}

To verify the practicability of ghost imaging for data imaging, we image an actual circuit signal with ghost imaging. The results are shown in Figure 3. Speckle pattern is directly converted from digital signal. The white speckle in the speckle pattern corresponds to the high level and the black speckle corresponds to the low level [Figs. 3 (a) and (c)]. When there  is noise in the actual circuit signal (the SNR is approximately 1:9), or when there is noise in the actual circuit signal (the SNR is approximately 1:2) and phase uncertainty (the random phase amplitude is approximately 290 degrees), ghost imaging can clearly show the periodicity of the data in the image. The image brightness distribution is consistent with that of ideal data.

Finally, we discuss the influence of noise in the experimental system on data ghost imaging. In theory, the random factors in the data, such as noise, random amplitudes and phase changes,
affect only the image corresponding to the object, while the nonrandom characteristics of the data are shown in the background area of the imaging,
 so the signal and noise can be perfectly separated. This means that in theory, any strong random noise does not affect the extraction of nonrandom characteristics from the data.
 However, the strong noise in the actual experiment affects the extraction of the data characteristics, mainly due to the additional noise generated during the experiment.
In the current experiment, we use a common commercial display, data acquisition card and barrel detector, which proves that this experiment is very easy to implement, but if one wants to further improve the anti-jamming ability of data ghost imaging, displays with higher contrast and faster response speed such as OLED displays, data acquisition cards with higher accuracy and low noise photoelectric detectors with low temperature thermostat systems can be used. In additional, When there are multiple periodic signals, similar to the discussion under Eq. (1), the resulting image will be the superposition of multiple periodic signals after removing noise. Further analysis of the image obtained by correlation reconstruction can obtain the frequency and amplitude of their respective periodic signals. Therefore, ghost imaging can also be used for the analysis of multiple periodic signals.

In summary, the periodicity of data is successfully imaged by using a computational ghost imaging experiment. It is found that the randomness factors in the data, such as noise,
random amplitudes and phase changes, mainly affect the image corresponding to the object, and the periodicity in the data is clearer in the background area of the image,
 so the signal and noise can be separated. The experimental results show that the periodicity in binarized data can be detected when the SNR is 1:19 or when the SNR is 1:7,
  the random amplitude of the phase reaches $210^\circ$, and the random amplitude reaches 90\%.
 This kind of data ghost imaging system is very simple and can be used in many fields such as big data analysis, meteorology, astronomy, and economics.

This work was supported by National Natural
Science Foundation of China (NSFC) (Grants
No. 11764008, No. 61774168, and  No. 11964007, the National Key
Research and Development Program of China (Grant
No. 2016YFA0202300), the Science and Technology Talent
Support Project of the Department of Education in
Guizhou Province (Grant No. KY[2018]045), the Science and Technology Foundation of Guizhou Province, China (Grant No.
[2020]1Y026), and
Construction project of characteristic key laboratory in Guizhou
Colleges and Universities (KY[2021]003).




\end{document}